# Hyperbolic absolute instruments

Tao Hou[1,2] and Huanyang Chen[1,*]

[1]*Department of Physics, Xiamen University; Xiamen, 361005, China*
[2]*Jiujiang Research Institute of Xiamen University; Jiujiang, 332000, China*
*\*kenyon@xmu.edu.cn*

**Abstract:** As a lens capable of sending images of deep sub-wavelength objects to the far field, the hyperlens has garnered significant attention for its super-resolution and magnification capabilities. However, traditional hyperlenses require extreme permittivity ratios and fail to achieve geometrically perfect imaging, significantly constraining their practical applications. In this paper, we introduce the general versions of hyperbolic absolute instruments from the perspective of dispersion and fundamental optical principles. These instruments enable the formation of closed orbits in geometric optics, allowing hyperlenses to achieve aberration-free, perfect imaging. This development not only provides a flexible and practical tool for enhancing the performance of traditional hyperlens, but also opens new possibilities for new optoelectronics applications based on hyperbolic ray dynamics.

## 1. Introduction

The resolution of a conventional optical system is limited to approximately half of the working wavelength [1, 2], which stems from that evanescent waves carrying the detailed information decays rapidly and cannot reach the far field. In recent years, various advanced imaging lenses have been proposed to overcome this resolution barrier by effectively manipulating evanescent waves. These imaging lenses are generally classified into two categories: flat lenses [3-11] and cylindrical lenses [12-18]. While flat lenses are more convenient for microscopy applications [19], they cannot transmit information to the far field without external assistance as the waves outside the lens remain evanescent. In contrast, cylindrical lenses can magnify images and address this limitation [20] by compressing the waves with high wavevectors to propagate to the far field as they move towards the outer edge of the cylindrical lens.

Here we primarily focus on two significant types of cylindrical imaging lens: absolute optical instrument (AI) [21] and hyperlens. AIs can provide a perfectly sharp image of all points in some spatial region without geometrical aberrations, but they are incapable of supporting the propagation of evanescent waves. Hyperlens [22, 23], on the other hand, has been proposed to convert the evanescent waves into propagating wave, enabling super-resolution imaging. However, hyperlens requires extreme permittivity ratios and cannot achieve perfect imaging, often resulting in severe caustics. Recent proposal of perfect hyperlens [24] underscores the potential of combining the advantages of hyperlens and perfect lens [3], yet it is still a flat lens. Thus, developing a method for designing AIs in hyperbolic systems is highly desirable for optical imaging systems but remains unexplored.

In this paper, we propose the concept and designing methodology of hyperbolic absolute instrument (HAI). By analyzing hyperbolic dispersion, we demonstrate that HAI can circumvent the topological singularity of traditional hyperlens and form closed orbits in geometric optics for hyperlenses. To further elucidate hyperbolic ray dynamics, we derive the angular momentum and turning parameter of radially hyperbolic anisotropic media (RHAM). Finally, we present the general forms of electromagnetic parameters for HAIs, and demonstrate several profiles with these intriguing properties. Compared to traditional AI with isotropic dispersion, HAI supports beamlike radiation patterns and offers multiple controlling degrees of freedom, enabling superior field regulation and enhanced focusing ability in wave optics.

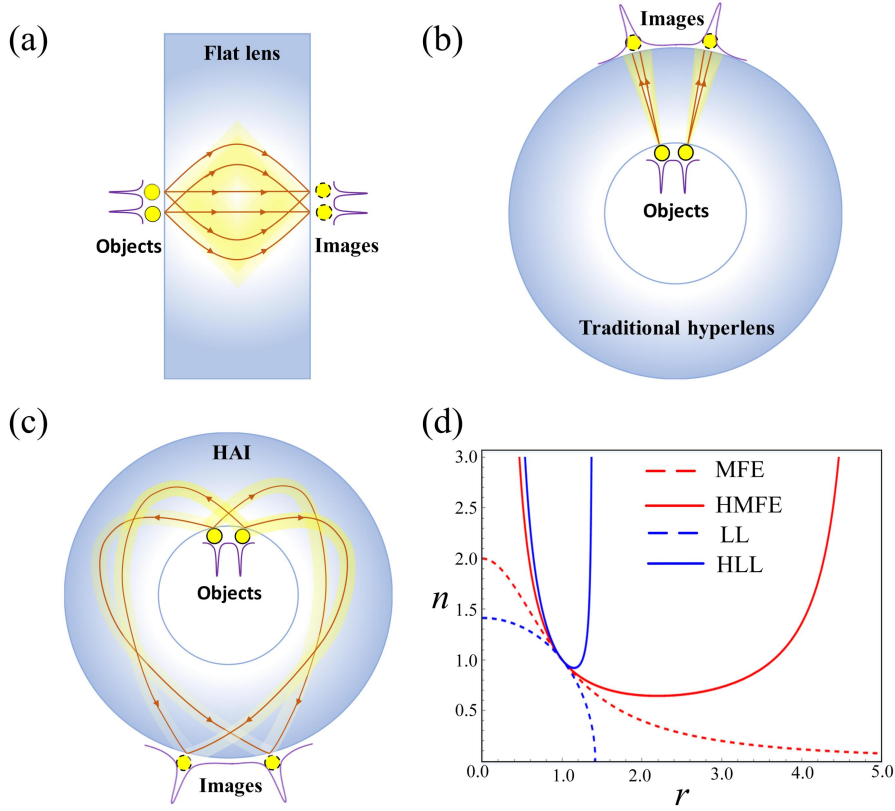

Figure 1 Schematic sketches of (a) the flat lens, (b) the traditional hyperlens, and (c) the HAI. Two objects in close are imaged perfectly by the action of different lenses. (d) The gradient refractive indexes $n$ as a function of radius $r$ for Maxwell's fish-eye (MFE), hyperbolic Maxwell's fish-eye (HMFE), Luneburg lens (LL) profile and hyperbolic Luneburg lens (HLL) profile.

## 2. Results and discussion

Firstly, we present the metric of perfect hyperlens in our recent work [24]

$$ds^2 = n^2(dx^2 - dy^2) = \frac{n_0^2}{\cos^2(y)}(dx^2 - dy^2). \tag{1}$$

Notably, this lens, inspired from the de Sitter spacetime in cosmology, inherits the self-focusing properties of Mikaelian lens [25] and achieves enhanced resolution through hyperbolic dispersion. However, it remains a flat lens and cannot sent information to the far field (see Fig.1(a)).

Furthermore, we apply an exponential conformal mapping $w=e^z$ [26], transforming the flat metric into a cylindrical one $ds^2 = n^2(-dr^2 + rd\varphi^2) = (\frac{1}{r\cos(\ln r)})^2(-dr^2 + rd\varphi^2)$. In the following discussions, we demonstrate that it can support closed orbits in geometric optics as the same as Maxwell's fish-eye (MFE) [27], a well-known AI with constant curvature. Beyond this specific example, more general absolute instruments do not necessarily exhibit constant curvature but can still support closed trajectories. This raises a fundamental question: How can we develop a systematic approach for designing such AIs with radially hyperbolic dispersions, i.e. HAIs?

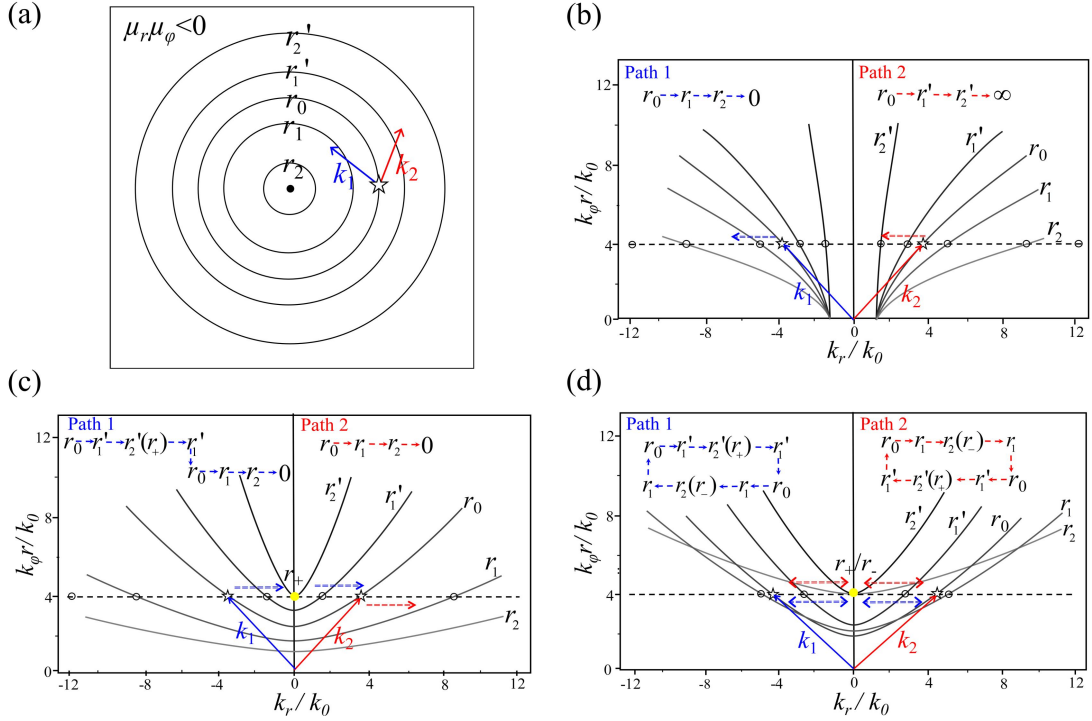

Figure 2 (a) Radially hyperbolic anisotropic space discretized based on the value of radius. The corresponding hyperbolic dispersion relations are illustrated for different cases: (b) $\mu_r < 0$, $\mu_\varphi > 0$, (c) $\mu_r > 0$, $\mu_\varphi < 0$, $\mu_r > 0$ with a constant refractive index $n = 1$, and (d) $\mu_\varphi < 0$, $\mu_r > 0$ with an ideally gradient refractive index profile $n(r)$. Here, we sketch five isofrequency contours with different radii as examples. The blue and red arrows denote the wave vectors $k_1$ ($k_r<0$) and $k_2$ ($k_r>0$), respectively. According to phase-matching conditions (black dotted lines), we can depict the evolutions of the wave vectors along path 1 (blue isometric arrows) and path 2 (red isometric arrows). Among them, the white stars, the white circles and yellow circles represent incident points, approach points and turning points, respectively.

## 2.1 Hyperbolic dispersion analysis

To begin, we focus on the RHAM where permeabilities exhibit opposite signs in the tangential and radial directions ($\mu_r < 0$, $\mu_\varphi > 0$ or $\mu_r > 0$, $\mu_\varphi < 0$). The corresponding dispersion relations [22] are

$$\frac{k_r^2}{\mu_\varphi} - \frac{k_\varphi^2}{|\mu_r|} = n^2 \frac{\omega^2}{c^2},$$
$$-\frac{k_r^2}{|\mu_\varphi|} + \frac{k_\varphi^2}{\mu_r} = n^2 \frac{\omega^2}{c^2}$$
(2)

where $k_r$ and $k_\varphi = l/r$ represent the radial and tangential components of the wave vector $k$, and $l$ is the angular momentum mode number. For convenience, we discretize the real space (see Fig. 2(a)) and plot the dispersion between $k_r$ and $k_\varphi r$ ($l$) at different radii in Figs. 2(b-d). Here we consider two incident beams with wave vectors $k_1$ ($k_r<0$) and $k_2$ ($k_r>0$). According to phase-matching conditions ($l_{in}=l_{out}=l_0$), the propagation of wave in real space corresponds to the transitions between different isofrequency contours. Notably, the propagation directions in anisotropic systems are determined by the Poynting vector rather than the wave vector.

Turning point [28] is a key physical quantity for analyzing the ray motions in AIs. In general, the centrifugal/centripetal rays will transition into centripetal/centrifugal rays at the turning points. Hence, the rays satisfy $dr/dt=0$ or $k_r=0$ at these points, i.e. the intersections of isofrequency contours with the $l$-axis. Previous works told that there are two distinct turning points ($r_+$ and $r_-$) or infinite turning points ($r_+ = r_-$) during the propagation of rays in the AIs. For simplicity, we first consider the simplest case of $n=1$. In the traditional hyperlens ($\mu_r < 0$, $\mu_\varphi > 0$), hyperbolic dispersion causes wave to either travel toward the center or diverge, as they cannot bypass the turning points for any $l_0$ (see Fig. 1(b) and Fig. 2(b)). Consequently, traditional hyperlenses fails to achieve perfect imaging. For the other case ($\mu_r > 0$, $\mu_\varphi < 0$) in Fig. 2(c), the beam with wave vector $k_1$ encounters the only turning point ($r_+$) once before traveling toward the center (path 1), while the beam with wave vector $k_2$ is eventually trapped at the center (path 2).

Obviously, this latter case aligns more closely with the design requirements of HAIs. Then we introduce an ideal refractive index profile $n(r) \neq 1$ for the case of ($\mu_r < 0$, $\mu_\varphi > 0$), generating two isofrequency contours ($r=r_\pm$) intersecting on the $l$-axis. Under these conditions, the beam with wave vector $k_1$ reaches the outer turning point $r_+$ before turning into the center (path 1), while the beam with wave vector $k_2$ reaches the inner turning point $r_-$ and turns away from the center (path 1). Eventually, the beams following the path 1 and path 2 reconverge at the incident point $r=r_0$, forming closed orbits (see Fig. 2(d)). Therefore, the above hypothetical profile meets our preview of HAIs (see Fig. 1(c)).

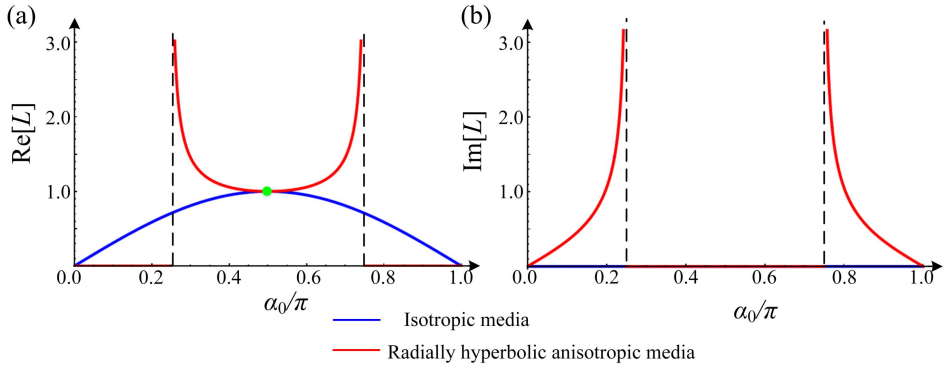

Figure 3 Relationship between the (a) real part and (b) imaginary part of the angular momentum $L$ and the incident angle $\alpha_0$. For convenience, the initial turning parameter is set as $N(r_0) = 1$. The blue curves represent the isotropic medium, while the red curves correspond to the RHAM. The green dot marks the intersection of the two curves at (0.5, 1). The dotted lines indicate the boundaries of the angular momentum.

### 2.2 Methods for designing HAIs

To derive the exact solution of the above refractive index, we start from a special hyperbolic line element with an assumed index $n(r)$

$$ds^2 = n^2(-\gamma^2 dr^2 + r^2 d\varphi^2) . \qquad (3)$$

Here we define a physical space ($r$, $\varphi$) and a virtual space ($r$, $\varphi'$), where the coordinate transformation between them satisfies $\varphi'=\varphi/i\gamma$, with $\gamma$ being a real and positive constant. Under this transformation, the line element becomes

$$ds^2 = -n^2\gamma^2(dr^2 + r^2 d\varphi'^2) = -n^2\gamma^2 dl^2, \qquad (4)$$

where $dl$ can be interpreted as the particle trajectory in virtual space. In physical space, the angular momentum of anisotropic system [29] can be written as

$$L = n^2 r^2 \frac{d\varphi}{ds}$$
$$= nr\frac{rd\varphi'}{dl} \qquad (5)$$
$$= nr\sin\alpha',$$

where $\alpha'$ represents the angle between the tangent to the particle trajectory and the radius vector in the virtual space. For a cylindrically symmetric medium, we introduce a turning parameter [28] defined as $N(r)=nr$. Considering a light ray propagating with angular momentum $L$ in the physical space, it follows from Eq. (5) that $L=N\sin\alpha'$ and the radius $r$ will reach the maximum or minimum at the turning points $r_\pm$, where $N(r=r_\pm)=L$. At a general point along a ray trajectory, the derivative of the polar angle $\varphi$ satisfies

$$\tan\alpha = \frac{rd\varphi}{dr} = i\gamma\frac{rd\varphi'}{dr} = i\gamma\tan\alpha' = \pm\frac{\gamma L}{\sqrt{L^2-(nr)^2}}, \qquad (6)$$

where $\alpha$ represents the angle between the tangent to the particle trajectory and the radius vector in the physical space. Using Eq. (6), we express $\alpha'$ as $\alpha'=\arctan(\tan\alpha/i\gamma)$. It is worth mentioning that for any free trajectory (i.e., without external force, collision, etc), the quantity $L$ remains invariant once the initial position $r_0$ and incident angle $\alpha_0$ are given. Consequently, the angular momentum of RHAM can also be written as

$$L = N(r_0)\sin\alpha_0 = N(r_0)\arctan\tan\alpha_0/i\gamma). \qquad (7)$$

In Fig. 3, we illustrate the relationship between angular momentum $L$ of RHAM and incident angle $\alpha_0$ for $\gamma=1$, with the isotropic cases provided for comparison. We find that the angular momentum of radially hyperbolic system takes the real value only when $\alpha_0 \in \{\pi/4, 3\pi/4\}$; otherwise, it takes the imaginary value. This indicates that light can propagate stably only for the incidence angle $\alpha_0$ within this range $\{\pi/4, 3\pi/4\}$ which closely corresponds to the opening angle of the hyperbolic dispersion discussed above. Furthermore, the angular momentums in RHAM and isotropic media (IM) always satisfy the relationship $L_{RHAM} \geq N(r_0) \geq L_{IM}$ with their curves intersecting at the special point $\alpha_0=\pi/2$. In other words, the incident angle in virtual space needs to satisfy a special condition $|\sin\alpha_0'|\geq 1$ to ensure ray propagation in physical space. Therefore, different from the isotropic case $L\leq N(r)$ discussed in previous studies [30], all the points along the propagation path need to satisfy the condition $L\geq N(r)$.

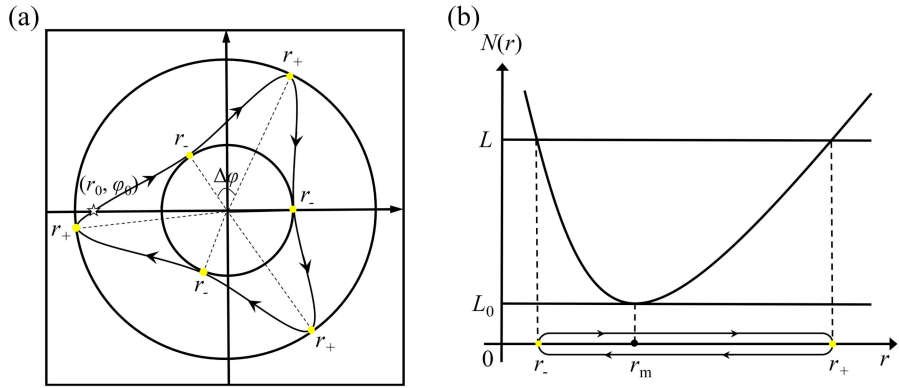

Figure 4 (a) Motion diagram of a ray in the HAI. (b) Relationship between the turning parameter $N(r)$ and radius $r$. Among them, the white star and the yellow circles denote the incident point and the turning points, respectively.

Considering a light ray originating from the origin $(r_0, \varphi_0)$, we obtain from Eq. (6) that

$$\varphi - \varphi_0 = \gamma L \int_{r_0}^{r} \frac{dr}{r\sqrt{L^2 - N(r)^2}}, \tag{8}$$

which serves as a fundamental equation of light ray in the RHAM with cylindrical symmetry. In AI, the ray follow a closed trajectories, oscillating between two circles of radii $r_-$ and $r_+$, touching them at equal angular intervals (see Fig. 4(a)). To ensure such a trajectory, we assume that the function $N(r)$ decreases for $r \leq r_m$ and increases for $r \geq r_m$ with the radius $r_m > 0$ (see Fig. 4(b)). Consequently, $N(r)$ attains a global minimum $L_0$ at the point $r = r_m$, i.e. $N(r_m) = L_0$. In order to transform Eq. (8) into a known integral equation, let us introduce the variable $\tau = \ln(r)$. During the motion between two turning points $r_-(L)$ and $r_+(L)$, the increment of the polar angle in physical space can be expressed as

$$\Delta\varphi(L) = \gamma L \int_{\tau_-}^{\tau_+} \frac{d\tau}{\sqrt{L^2 - N(\tau)^2}}. \tag{9}$$

Here we set $\Delta\varphi = \pi/m' = \gamma\pi/m$, where $m'$ is a rational number ensuring the closed rays in HAI. In Note S1 of Supplemental materials, we solve this integral equation of Abel's type and obtain the general index profile as

$$n(r) = \frac{L_0}{r\cos(\frac{m}{2}\ln(f(r)/r))}, \tag{10}$$

Here $f(f(r)) = r$ and graph of $f(r)$ is symmetric with respect to the axis $f(r)=r$. By substituting Eq. (10) into Eqs. (3) and (5), we obtain the complete expressions for angular momentum and line element

$$L = nr\sin\alpha' = \frac{L_0 \sin\alpha'}{\cos(\frac{m}{2}\ln(f(r)/r))}$$

$$ds^2 = \left(\frac{L_0}{r\cos(\frac{m}{2}\ln(f(r)/r))}\right)^2 (-\gamma^2 dr^2 + r^2 d\varphi^2). \tag{11}$$

When $\gamma$ and $m$ are set as $i$, Eq. (11) will degenerate into the case of traditional AIs [28]. Based on transformation optics, the permittivity and permeability of Eq. (9) can be expressed as

$$\varepsilon^{ij} = \mu^{ij} = \pm\sqrt{|g|}g^{ij} = \pm\begin{bmatrix} -1/\gamma & & \\ & \gamma & \\ & & \gamma n^2 \end{bmatrix} \tag{12}$$

Considering the 2D TE polarization and the dispersion requirement ($\mu_r > 0$, $\mu_\varphi < 0$), we select the dominant parameters $\{\mu_r, \mu_\varphi, \varepsilon_z\} = \{1, -\gamma^2, n^2\}$ to screen out the self-focusing waves in wave optics. To simplify the complexity of the permittivity, an alternative selection $\{\mu_r, \mu_\varphi, \varepsilon_z\} = \{n^2, -\gamma^2 n^2, 1\}$ achieves the same effect. Similarly, the dominant parameters in 2D TM polarization can be written by replacing the position of permittivity and permeability. With these formulations, we establish a framework for HAI design. Furthermore, our approaches extend naturally to corresponding flat hyperlens by a logarithmic conformal mapping $z=\ln w$, finding exciting applications in cosmology [31] and polariton regulation [24]. Their metric and the corresponding dominant parameters of 2D TE polarization can be written as

$$ds^2 = n_f^2(dx^2 - \gamma^2 dy^2) = \left(\frac{L_0}{\cos(\frac{m}{2}(\ln(f(\exp(y))) - y))}\right)^2 (dx^2 - \gamma^2 dy^2), \tag{13}$$

and $\{\mu_x, \mu_y, \varepsilon_z\} = \{-1, \gamma^2, n_f^2\}$ or $\{\mu_x, \mu_y, \varepsilon_z\} = \{-n_f^2, \gamma^2 n_f^2, 1\}$.

Notably, there are beamlike radiation patterns and well-defined boundaries in HAI systems, strictly constraining the incident direction and propagation regions. In geometrical optics, the incident directions need to be satisfied $[\frac{dr}{rd\varphi}]_{initial} < \sqrt{|\frac{\mu_r}{\mu_\varphi}|} = \frac{1}{\gamma}$, which can be derived by Eq. (6) and the divergence angle of hyperbolic dispersion $\theta_c$=arctan ($\sqrt{|\mu_r/\mu_\varphi|}$) [32]. Moreover, in order to avoid singularity $n=\infty$, the radius $r$ must satisfy the conditions $r\neq 0$ and $f(r)/r \neq \exp(m(p\pi+\pi/2)/2)$ ($p$ is an arbitrary integer), leading to periodic limiting boundaries in the ray propagation. Interestingly, the allowed range of the incident direction only depends on the value of $\gamma$, regardless of $m$ and $f(r)$, while the positions of the boundaries only depends on the value of $f(r)$ and $m$, independent of $\gamma$. In the following section, we will demonstrate the designing methodology and some unique characteristics of HAIs by discussing several famous examples. For convenience, we set $L_0=1$ in the subsequent discussions.

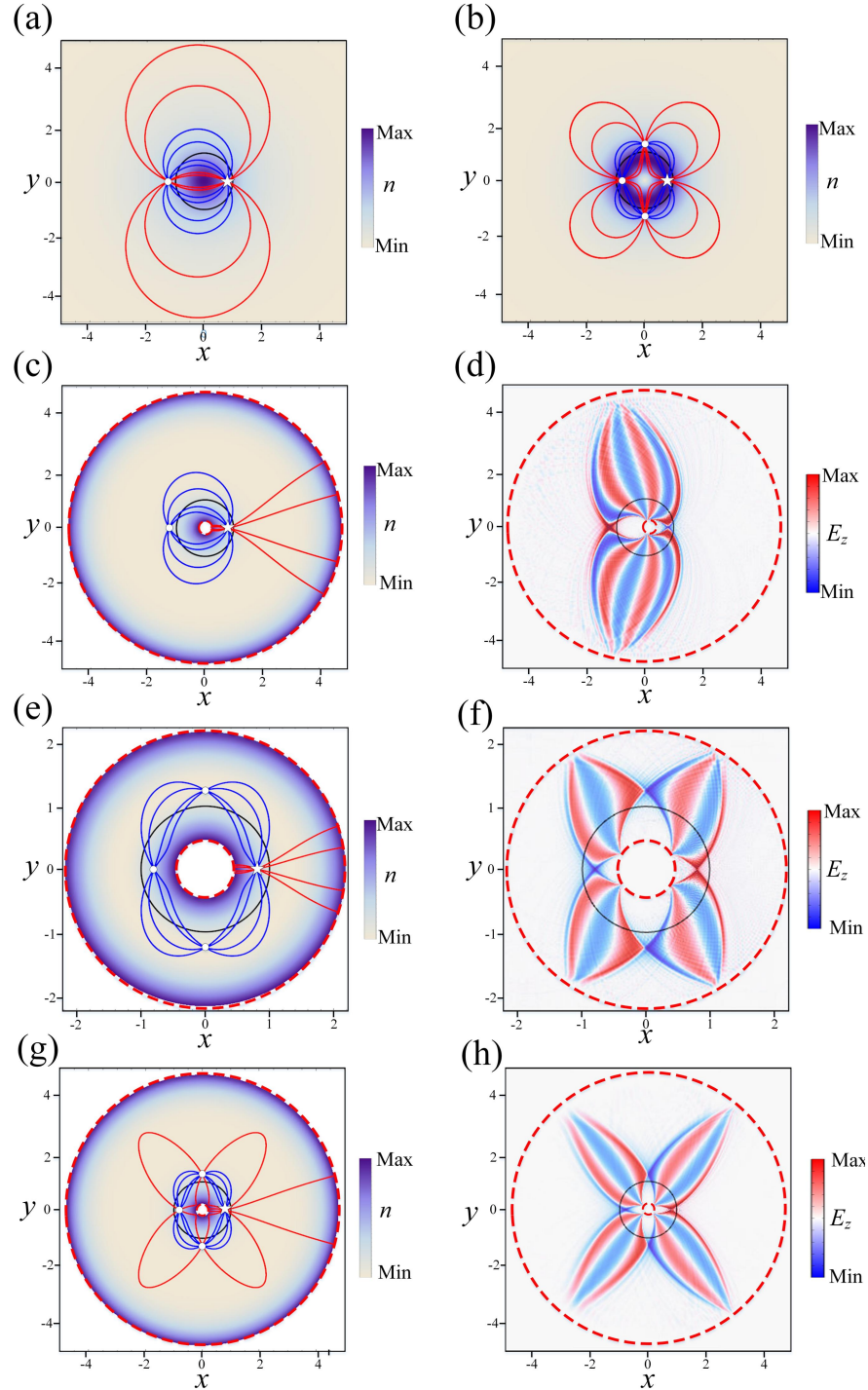

Figure 5 The rays and electric field $E_z$ in the (a) MFE, (b) GMFE, (c-d) HMFE, (e-f) HGMFE I [$m$=2, $\gamma$=1] and (g-h) HGMFE II [$m$=1, $\gamma$=0.5]. The white stars and circles represent the incident points ($r_0$, $\varphi_0$) = (0.8, 0) and the images respectively. The red and blue curves represent the rays with incident direction $\left|\frac{dr}{rd\varphi}\right|_{initial} > 1$ and $\left|\frac{dr}{rd\varphi}\right|_{initial} < 1$ respectively. The black solid boundaries and red dotted boundaries represent the condition of $n$=1 and $n$=∞, respectively.

### *2.3 Examples of HAIs*

Above all, we consider the case of traditional MFE, where the refractive index profile is given by $n$=1/($r^2$+1). In this case, the object position $r_A$ and the image position $r_B$ are also turning points, satisfying the relation $r_A r_B$ =1 (see Fig. 5(a)). To model this in the hyperbolic framework, we set $f(r)$ = 1/$r$ and $m$=1 for the case of $\gamma$=1. This leads to the gradient refractive index of hyperbolic Maxwell's fish-eye (HMFE) $n$=1/cos(In($r$)). Interestingly, it is also the cylindrical analog of the perfect hyperlens [24], confirming the universality of our method. With the increasing of radius $r$, the index function of HMFE follows a parabolic variation, in contrast to the monotonically decreasing profile of the traditional MFE lens (see Fig. 1(d)). For demonstration, we take 2D TE mode for example, where the dominant parameter of HMFE are given by {$\mu_r$, $\mu_\varphi$, $\varepsilon_z$}={1, -$\gamma^2$, (1/cos(In($r$)))$^2$}. To verify its geometric effect, we set the incident point ($r_0$, $\varphi_0$) = (0.8, 0) to launch rays along different directions. Among them, The red and blue curves represent the rays propagating along the incident directions $\left|\frac{dr}{rd\varphi}\right|_{initial} > 1$ and $\left|\frac{dr}{rd\varphi}\right|_{initial} < 1$, respectively. In wave simulation, a line source along the $z$-axis is placed at the same position ($r_0$, $\varphi_0$) = (0.8, 0) to stimulate the field. Due to the beamlike radiation patterns and boundaries in hyperbolic systems, some rays from the object ($r$, $\varphi$) focus on the images ($r^{-1}$, $\varphi+\pi$) and return back to the origin, while others diverge towards the boundaries (see Fig. 5(c)). Nevertheless, we can observe perfect hyperbolic focusing wavefront in Fig. 5(d).

In the above analysis, HAI provides an additional degree of freedom $\gamma$ to control light compared with traditional AI. To further understand the effects of $\gamma$ and $m$, we show two types of HGMFE (I[$m$=2, $\gamma$=1] and II[$m$=1, $\gamma$=0.5]) with the same $\Delta\varphi$ in Figs. 5(e-h). Notably, there are three images at the same positions in HGMFE as in the traditional generalized Maxwell's fish eye (GMFE) in Figs. 5(b). However, the type II support the propagating wave with a broader beam range and wider boundary compared to type I influenced by the value of $m$ and $\gamma$. This demonstrates that the flexible control of $m$ and $\gamma$ can greatly enhance the design and application potential of HAIs in future.

In Note S2 of Supplemental materials, we also explore other HAIs that focuses all the rays from the object $r_A$ to the image $r_B$ with the same radius or back to the object, which corresponding to the Luneburg lens profile and Eaton/Miñano lens profile.

### 3. Conclusion

In this paper, we extend the concept of traditional absolute optical instrument into the hyperbolic domain by introducing the general forms of HAI and the corresponding flat hyperlens. We demonstrate that HAI retains the geometrically perfect imaging property of traditional AI while uniquely enabling the propagation of evanescent wave with high wave vector. This capability paves the ways for numerous applications, such as super-resolution real-time imaging, sensing, and so on. Although the general electromagnetic forms of HAIs are complex, their realization may be feasible in the future through carefully manipulating the thickness of two-dimensional materials [33], the filling factor of concentric multilayer metamaterials [34] or the geometry of split-ring resonators [35].

Previous studies suggest that incident beams in the RHAM are attracted towards the origin along spirallike trajectories due to topological singularity [32, 34, 36]. Notably, our work circumvents the problem by employing special refractive indexes and derives the propagation processes of light rays in RHAM with multiple degrees of freedom, which are expected to stir up great interest in exploring more intriguing hyperbolic devices with novel functionalities, such as revolution, magnification, invisibility, and so on. Looking ahead, we anticipate the discovery of new classes of AIs with complex electromagnetic parameters, potentially capable of self-focusing on complex manifold, thereby opening exciting avenues for imaging in non-Hermitian systems.

**Funding.** This research was supported by the National Key Research and Development Program of China (Grant No. 2023YFA1407100 and No.2020YFA0710100), National Natural Science Foundation of China (Grants No.12361161667), and Jiangxi Provincial Natural Science Foundation (Grant No.20224ACB201005).

**Disclosures.** The authors declare no conflicts of interest.

**Data availability.** Data underlying the results presented in this paper are not publicly available at this time but may be obtained from the authors upon reasonable request.

**Supplementary document.** See Supplemental materials for supporting content.

# Supplemental Materials for

# "Hyperbolic absolute instruments"

Tao Hou[1,2] and Huanyang Chen[1,*]

[1]*Department of Physics, Xiamen University; Xiamen, 361005, China*
[2]*Jiujiang Research Institute of Xiamen University; Jiujiang, 332000, China*
*\*kenyon@xmu.edu.cn*

**Note S1: The calculation of integral in Eq. (9).**

**Note S2: Other examples of hyperbolic absolute instruments.**

**Figs. S1 – S2**

**Supplementary Reference**

**Note S1: The calculation of integral in Eq. (9).**

In order to solve the integral equation of Abel's type in Eq. (9), we divide $\Delta\varphi(L)$ by $\gamma\sqrt{L^2 - L'^2}$, where $L'$ is an integration parameter, and integrate with respect to $L'$ from $L_0$ to $L$:

$$\int_{L_0}^{L} \frac{\Delta\varphi(L')dL'}{\gamma\sqrt{L^2 - L'^2}} = \int_{L_0}^{L} \int_{\tau_-(L')}^{\tau_+(L')} \frac{d\tau}{\sqrt{L'^2 - N(\tau)^2}} \frac{L'dL'}{\sqrt{L^2 - L'^2}}$$

$$= \int_{\tau_-(L)}^{\tau_+(L)} \int_{N(\tau)}^{L} \frac{L'dL'}{\sqrt{L^2 - L'^2}} \frac{d\tau}{\sqrt{L'^2 - N(\tau)^2}}$$

$$= \int_{\tau_-(L)}^{\tau_+(L)} \left[ -\arcsin\sqrt{\frac{L^2 - L'^2}{L^2 - N(\tau)^2}} \right]_{L'=N(\tau)}^{L'=L} d\tau \quad (S1)$$

$$= \int_{\tau_-}^{\tau_+} (0 - (-\frac{\pi}{2})) d\tau = \frac{\pi}{2}(\tau_+ - \tau_-)$$

where we have inverted the order of integration and adjusted the integration limits appropriately (see Fig. S1) [28]. Here, we set $\Delta\varphi = \pi/m' = \gamma\pi/m$, where $m'$ is set as a rational number to satisfy the closed ray in HAI. Then we can obtain from Eq. (S1):

$$\ln\frac{r_+}{r_-} = \frac{2}{m}\arccos(L_0/L). \quad (S2)$$

Here we define a function $f(r)$, and it satisfy the conditions $f(r_-) = r_+$ and $f(r_+) = r_-$ for arbitrary turning points $r_\pm$. i.e.

$$r_\pm = f(r_\mp). \quad (S3)$$

Then we obtain $f(f(r)) = r$. This implies that the graph of $f(r)$ is symmetric with respect to the axis $f(r) = r$ and intersects at the point $r = r_m$ when satisfying the condition $r_+ = r_-$. Notably, the intersection corresponds to the case $L = L_0$ (the lowest point of the graph in Fig. 4(b)) i.e. circular ray trajectory. With the function $f(r)$ defined this way, we can express $r_-$ in terms of $f(r_+)$ or vice versa. Taking the fact that $L = N = nr$ at $r = r_\pm$ and omitting the index of $r_\pm$, we obtain the general index as

$$n(r) = \frac{L_0}{r\cos(\frac{m}{2}\ln(f(r)/r))}. \quad (S4)$$

**Note S2: Other examples of hyperbolic absolute instruments.**

*Hyperbolic Luneburg lens (HLL) profile*

As a well-known AI, Luneburg lens (LL) profile ($n=\sqrt{(2-r^2)}$) focuses all the rays from the object $r_A$ to the image with the same radius $r_B = r_A$, and the corresponding ray trajectories form ellipses centered at the origin [S1] (see Fig. S2(a)). Here, we adopt the same $f(r) = \sqrt{(2-r^2)}$ and $m=2$ as LL for the case of $\gamma=1$. The ray and wave in HLL still retain the focusing characteristic of LL profile even though the refractive index distribution is completely different (see Fig. 1(d) and Fig. S2(b-c)). While both Luneburg lens profile and HLL profile have similar outer boundary, parts of rays can bypass the former but diverge to the latter. It is worth noting that there is interference in areas other than the vicinity of the focal point due to the convergence of multiple rays during propagation.

*Hyperbolic Eaton/Miñano lens (HEL) profile*

Traditional Eaton/Miñano lens (EL) profile ($n=\sqrt{(2/r-1)}$) corresponds to elliptic motion in the Newton potential in mechanics, where ray trajectories are confocal ellipses with focus at the origin and with the main semiaxes of unit length (see Fig. S2(d)) [S1]. Here we adopt the same $f(r)=2-r$ and $m=1$ for the case of $\gamma=1$, the ray and wave maintain self-focusing properties at the objects after rotating around the center (see Figs. S2(e-f)). Similar to HLL, interference effects arise due to the convergence of multiple rays during propagation.

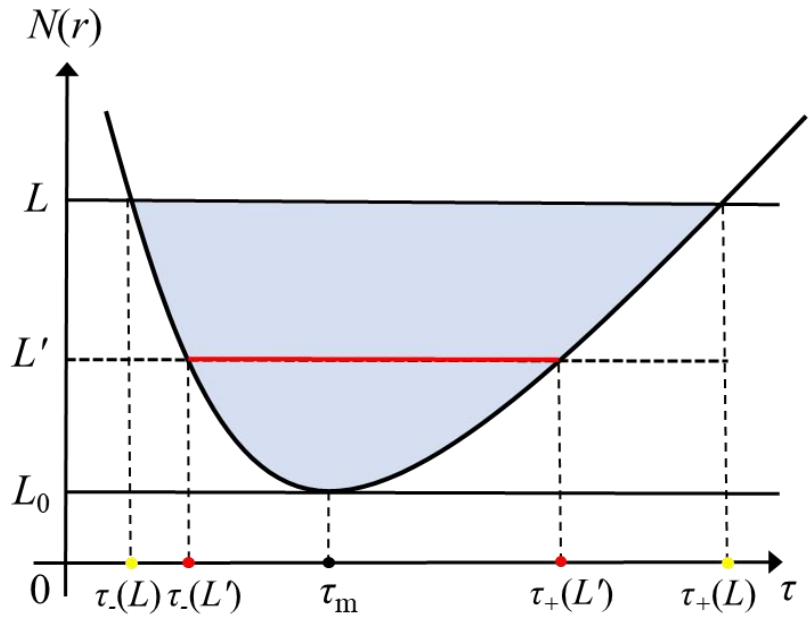

Figure S1 The change of integration limits in Eq. (S1) illustrating that
$\int_{L_0}^{L}\int_{\tau_-(L')}^{\tau_+(L')}\frac{d\tau}{\sqrt{L'^2-N(\tau)^2}}\frac{L'dL'}{\sqrt{L^2-L'^2}}=\int_{\tau_-(L)}^{\tau_+(L)}\int_{N(\tau)}^{L}\frac{L'dL'}{\sqrt{L^2-L'^2}}\frac{d\tau}{\sqrt{L'^2-N(\tau)^2}}$.

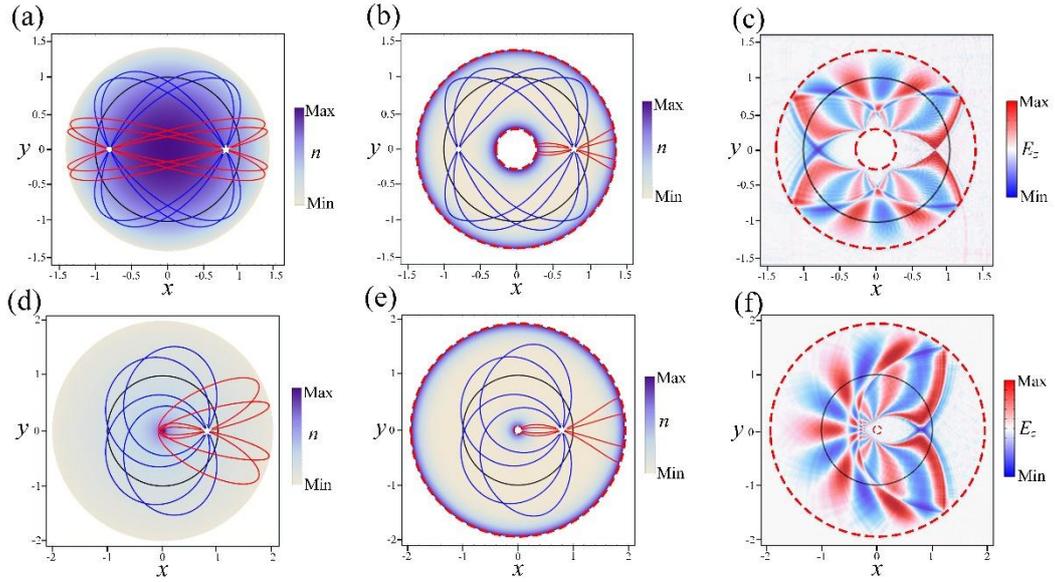

Figure S2 The rays and electric field $E_z$ in the (a) LL profile, (b-c) HLL profile, (d) EL profile and (e-f) HEL profile. The white stars and circles represent the incident points $(r_0, \varphi_0) = (0.8, 0)$ and the images, respectively. The red and blue rays represent the rays with incident direction $\left|\frac{dr}{rd\varphi}\right|_{initial} > 1$ and $\left|\frac{dr}{rd\varphi}\right|_{initial} < 1$, respectively. The black solid boundaries and red dotted boundaries represent the condition $n=1$ and $n=\infty$, respectively.

## Supplementary Reference

[S1] U. Leonhardt, and T. Philbin, *Geometry and Light: The Science of Invisibility* (New York: Dover, 2010)